\documentclass{iau_mod}

\title[Recoiling black holes] 
{Recoiling black holes: electromagnetic signatures, candidates, and astrophysical implications}

\author[S. Komossa]   
{S. Komossa}

\affiliation{
Technische Universit\"at M\"unchen, Fakult\"at f\"ur
Physik,\\ James-Franck-Strasse 1/I, 85748 Garching, Germany \\
Excellence Cluster Universe, TUM, \\ Boltzmannstrasse 2,
85748 Garching, Germany \\ 
Max-Planck-Institut f\"ur Plasmaphysik,\\ Boltzmannstrasse 2,
85748 Garching, Germany} 

\jname{Advances in Astronomy 2012}
\editors{doi: 10.1155/2012/364973}
\begin{document}

\maketitle

\begin{abstract}
Supermassive black holes (SMBHs) may not always
reside right at the centers of their host galaxies.
This is a prediction of numerical relativity simulations, which imply that
the newly formed single SMBH, after binary coalescence
in a galaxy merger,
can receive kick velocities up to several 1000 km/s
due to anisotropic emission of gravitational waves.
Long-lived oscillations
of the SMBHs in galaxy cores, and in rare cases even SMBH ejections
from their host galaxies, are the consequence.
Observationally, accreting
recoiling SMBHs would appear as quasars spatially and/or
kinematically off-set from their host galaxies.
The presence of the ``kicks'' 
has a wide range of astrophysical implications
which only now are beginning to be explored,
including consequences for black hole and galaxy assembly
at the epoch of structure formation, black hole feeding,
and unified models of Active Galactic Nuclei (AGN).
Here, we review the observational
signatures of recoiling SMBHs and the properties of the first candidates
which have emerged, including
follow-up studies
of the candidate recoiling SMBH of SDSSJ092712.65+294344.0.
\end{abstract}

\firstsection 

\section{Introduction}

Interaction and merging of galaxies occurs frequently throughout the
history of the universe. If both galaxies do harbor SMBHs, binaries
will inevitably form (Begelman et al. 1980). Galaxy mergers are believed
to be the sites of major black hole growth, and an active search for
SMBH pairs and binaries of wide and small separations is currently
ongoing (see Komossa 2006 for a review of electromagnetic signatures).    
When the two SMBHs ultimately coalesce, they are a source of strong gravitational
waves. These are emitted anisotropically during coalescence and carry
away linear momentum (e.g., Bekenstein 1973). As a result,  
the newly formed single SMBH recoils.
Configurations of coalescing black holes can lead to
kick velocities up to several thousand km/s
(e.g., Campanelli et al. 2007 and 2009, Gonz{\'a}lez et al. 2007 and 2009,
Herrmann et al. 2007, Baker et al. 2008,
Br\"ugmann et al. 2008, Dain et al. 2008, Miller \& Matzner 2009,
Lousto \& Zlochower 2009, Le Tiec et al. 2009, Lousto et al. 2010, Lousto \& Zlochower 2011a; 
review by Centrella et al. 2010). In the initial computations,
kick velocity was highest for maximally spinning equal-mass black hole 
binaries with anti-aligned spins
in the orbital plane (``superkicks'').  More recently, based on a new recoil
formula, Lousto \& Zlochower (2011b) have estimated that
recoil velocities up to 5000 km/s can be reached in configurations
with spins partially aligned with the orbital angular momentum.   
In unbound encounters (not likely to occur in astrophysical
environments), the kick velocity can exceed 15\,000 km/s 
(Healy et al. 2009, Sperhake et al. 2011).  

After the kick, the recoiling SMBH
will oscillate about the core of its 
host galaxy (Madau \& Quataert 2004, Gualandris \& Merritt
2008)
or will even escape, if
its kick velocity exceeds the escape velocity of its host.
In a ``typical'', gas-poor galaxy, a black hole kick velocity of 500 km/s
will result in an initial amplitude of $\sim$200 pc, and an oscillation
timescale of order 10$^7$ yrs (Fig. 1 of Komossa \& Merritt 2008b). 
The kicks, including those large enough to remove SMBHs from their host galaxies,
have potentially far-reaching astrophysical consequences,
including for SMBH and galaxy assembly and AGN statistics.
Upon recoil, the most tightly bound gas will remain bound to the recoiling
black hole, and therefore
high-velocity kicks imply the existence of interstellar and intergalactic quasars
(e.g.,  Madau et al. 2004, Merritt et al. 2004, Madau \& Quataert 2004,
Libeskind et al. 2006, 
Loeb 2007,
Gualandris \& Merritt 2008, Blecha \& Loeb 2008, Komossa \& Merritt 2008b, 
Volonteri \& Madau 2008, Liu et al. 2011).  
Identifying recoiling SMBHs through observations is of great interest.
Several key electromagnetic signatures of kicks have been
predicted in the last few years, and first candidate recoiling SMBHs have emerged.

This chapter is structured as follows. In Section 2, an overview of
the predicted electromagnetic signatures of recoiling SMBHs is given.
In Sect. 3 the event frequency is discussed,
while  Sects 4 and 5 provides
a review of the published candidate recoiling SMBHs. 
Sect. 6 explores
consequences of recoil for unified models of AGN.
Sect. 7 concludes
with some astrophysical consequences and important future studies.

\section{Electromagnetic signatures of recoiling SMBHs}

\subsection{Broad emission-line shifts}

After the kick, matter remains bound to the recoiling SMBH
within a region whose radius $r_{\rm k}$ is given by

$r_{\rm k} = {GM_{\rm BH}\over{v_{\rm k}^2}} \approx 0.4
 \left({M_{\rm BH}\over{10^8 {\rm M}_\odot}}\right)
\left({v_{\rm k}\over{10^3 {\rm km\ s}^{-1}}}\right)^{-2} ~{\rm pc},$

where $v_{\rm k}$ is the kick velocity (Merritt et al. 2006).
This region is on the order of the size of the broad line region (BLR) of AGN
(Peterson 2007). The accretion disk and BLR will therefore typically remain bound to the SMBH
while the bulk of the host galaxy's narrow-line region (NLR)
will remain behind.
The accreting recoiling SMBH will therefore appear as an off-nuclear ``quasar''
as long as its accretion supply lasts. However, spatial off-sets are challenging to
detect even with the {\it{Hubble Space Telescope}} (HST) except in the nearby universe.
The kinematic Doppler shifts of the broad emission lines are, in principle,
easy to measure out to high redshifts.
Spectroscopically,
 recoiling SMBHs will appear as AGN which have
their broad emission lines kinematically shifted by up 
to $\sim 5000$ km/s with respect to their NLRs.

Bonning et al. (2007) suggested several criteria, how to identify a recoiling SMBH
spectroscopically. Apart from (1) the kinematic shift of the BLR, a candidate
recoiling SMBH should (2) show symmetric broad line profiles, it should (3)
lack an ionization stratification of its narrow emission lines,
and it should (4) not show any shift between broad MgII and the broad Balmer
lines{\footnote{In practice, individual recoil candidates may show some (temporary) deviations
from this scheme, or exhibit extra features.  For instance,
just after recoil, the BLR emission profiles would likely be asymmetric.
Feedback trails from partially bound gas and disk winds would produce
emission-line signatures at various kinematic shifts between zero and the
recoil velocity. Once the SMBH has travelled beyond the extent of the classical NLR of
a few kpc extent, low-density ``halo'' gas would dominate the optical
narrow-line spectrum, with emission-line ratios characteristically
different from the classical NLR.}}.
One object, the quasar SDSSJ092712.65+294344.0,
fulfills all of these four
criteria and is therefore an excellent candidate
for a recoiling SMBH (Komossa et al. 2008).
It will be further discussed in Section 4, together with several other candidate recoiling
BHs.
More candidates may hide in large samples of peculiar broad-line emitters recently
identified in the Sloan Digital Sky Survey (SDSS; Eracleous et al. 2011).

\subsection{Flaring accretion disks}

In gas-rich mergers, an accretion disk is likely present,
even though the inner part of the disk may only
re-form after binary coalescence (Liu et al. 2003,
Milosavljevi{\'c} \& Phinney 2005, Loeb 2007, Ponce et al. 2011).
UV, soft X-ray, and IR flares could result from
shocks in the accretion disk surrounding the SMBH
just after recoil, or when the inner disk reforms
(e.g., Lippai et al. 2008, Shields \& Bonning 2008,
Schnittman \& Krolik 2008, Megevand et al. 2009, 
Rossi et al. 2010, Corrales et al. 2010, Tanaka et al. 2010, Zanotti et al. 2011,
Ponce et al. 2011).
These flares may last $\approx$10$^{4}$ yrs
and may be detectable in current and future sky surveys.

\subsection{Tidal disruption flares from stars around recoiling SMBHs}

Even in the absence of an accretion disk,
ejected SMBHs  will always carry a retinue of bound stars.
Observable effects related to these stars are therefore perhaps
the most universal signature of recoil.  
As the SMBH moves through the galaxy, the bound, and unbound, 
stars are subject to tidal disruption, leading to powerful X-ray flares
of quasar-like luminosity (Komossa \& Bade 1999, Bloom et al. 2011),
which would appear off-nuclear or even
intergalactic.  Komossa \& Merritt (2008a, KM08 hereafter)
computed disruption rates for the bound, and the unbound,
stellar populations under recoil conditions. 
In the resonant relaxation regime, they showed that
the rates are of order $10^{-6}$ yr$^{-1}$ for a typical
postmerger galaxy (Fig. 2 of KM08);
smaller than, but comparable to, rates for non-recoiling SMBHs.
At an early phase of recoil, the tidal disruption rate can
be much higher, when the SMBH experiences a full loss-cone, and travels
through the clumpy core environment of a recent merger (KM08).
The flare rate may temporarily reach values as high (Stone \& Loeb 2011) as during
the peak of the pre-merger binary phase (Chen et al. 2009). 

Another signature related to the stars bound to the recoiling SMBH
is episodic X-ray emission
from accretion
due to stellar mass loss. Mass loss provides a reservoir of
gas, and therefore also {\em optical emission lines from gas at the
recoil velocity} even in the initial absence of
a gaseous accretion disk.
Other consequences include
the presence of
intergalactic planetary nebulae and supernovae, after the
ejected SMBH has left its host galaxy (KM08).

All these signals would generically
be associated with recoiling SMBHs, whether or not the galaxy
merger is gas-rich or dry, and whether or not an accretion
disk is present initially, and
they would continue episodically for a time of $\sim 10$ Gyr (KM08).

\subsection{Hypercompact stellar systems}

While the ``tidal recoil flares'' are very luminous and can be detected
out to very large distances, the compact system of bound stars itself
will be detectable in the nearby universe, and would
resemble a globular cluster in total luminosity,
but with a much greater velocity dispersion due to the
large binding mass $M_{\rm BH}$ (Komossa \& Merritt 2008a).
Merritt et al. (2009) worked out the properties of these ``hypercompact stellar systems''
(HCSS), and related the structural
properties (mass, size, and density profile) of HCSSs to the
properties of their host galaxies
and to the amplitude of the kick.
Since the kick velocity is encoded in the
velocity dispersion of the bound stars, future
detection of large samples of HCSSs would therefore allow
us to determine empirically the kick distribution,
and therefore the merger history of galaxies in clusters.
Nearby clusters of galaxies are best suited to search for and identify HCSSs, and
$\sim$100 of them should be detectable within 2 Mpc of the center of the Virgo cluster
(Merritt et al. 2009).
Depending on the merger history of our Milky Way (O'Leary \& Loeb 2009), 
and the merger history of black holes in its globular
clusters (Holley-Bockelmann et al. 2008), 
100s of
low-mass HCSSs and rogue black holes may reside in the halo of our Milky Way, 
and a search for them is underway (O'Leary \& Loeb 2011).

\subsection{Other observable effects of recoil}

During the long-lived ``Phase II'' recoil oscillations (Gualandris \& Merritt 2008),
when the SMBH oscillation amplitude is on the torus scale,
the SMBH might efficiently accrete from the dense molecular gas
at {\em each} turning point, causing repeated flares of radiation
(Komossa \& Merritt 2008b).
Such flares would locally destroy the dust, while photoionization of
the dense surrounding gas would produce a strong emission-line response.
Such a signal would not only help in identifying kicks but could also be used
as a new probe of the properties of the torus itself.

Other signatures of recoiling SMBHs include
effects on the morphology and dynamics of the gaseous disk
of the host galaxy (Kornreich \& Lovelace 2008),
their imprints on the
hot gas in early-type galaxies (Devecchi et al. 2009), accretion from
the ISM (Fujita 2009), the possibility of star formation in the wake of
the SMBH trajectory (de la Fuente Marcos \& de la Fuente Marcos 2008),
and their influence on the jet structures in radio galaxies (Liu et al. 2011).

\section{The frequency of recoiling SMBHs in astrophysical environments}

Several factors affect the distribution of SMBH kick velocities and their
observability; the system parameters of the SMBH binary on the one hand
(mass ratio, spin magnitudes and spin directions), and the astrophysical
environment on the other hand.  

The frequency of high-velocity kicks depends on the distribution of mass ratios
and spins of the binary
SMBHs. In case of random distributions of spin directions, as expected in
gas-poor galaxy mergers, the
kick formula (e.g., Campanelli et al. 2007, 
Baker et al. 2008, Lousto \& Zlochower 2009)
has been used to predict the kick fraction in dependence of recoil
velocity (Campanelli et al. 2007, Schnittman \& Buonanno 2007,
Baker et al. 2008, Komossa \& Merritt 2008b).
In this case, kicks with velocities larger
than 500 km/s are relatively common 
(Fig. 1 of Komossa \& Merritt 2008b).
Spin precession further has the consequence that large kicks are deboosted
if the angle $\Theta$ between the spin of 
the more massive BH and the orbital angular momentum is initially small, 
while large kicks are boosted, if $\Theta$ is initially large
(Kesden et al. 2010).   

The other key factor is the astrophysical environment, which
determines the spin magnitude (set by the mechanism of SMBH mass growth
via random accretion, ordered accretion, or BH-BH merging; Volenteri et al. 2007)
and the timescale of spin alignment with the orbital angular momentum   
(e.g., Scheuer \& Feiler 1996, Natarajan \& Armitage 1999,
Bogdanovi{\'c} et al. 2007) in gas-rich galaxy mergers. 
The latter depends on the rapidity of
binary coalescence, the amount of gas accretion before versus after coalescence,
the accretion rate, the disk properties (e.g., the viscosity law across the
disk), and the mass of the SMBH.  
While the most massive black holes are more resistent to alignment,
the process is generally relatively fast in gas rich environments
(timescales of 10$^{5}$--10$^{9}$ yrs or less; Perego et al. 2009, Dotti et al. 2010). 

While initial results from numerical relativity have indicated
that kick velocities are low in this case, the whole parameter 
space is still being explored, and 
Lousto \& Zlochower (2011b) have recently shown that 
kick velocities up to 5000 km/s can be reached in configurations
with spins partially aligned with the orbital angular momentum.
As a consequence, the likelihood of observing high-velocity
recoils in gas-rich environments is significantly higher 
than in some previous estimates (their Fig. 3). 

Given the  large number of uncertain parameters in estimating the 
frequency of recoiling SMBHs, identifying
them  through observations is also important.  
Ultimately, observations will independently provide the distribution of kick velocities
and the properties of the recoiling SMBHs. First candidates have emerged in recent years,
and more are likely to come soon, given the number of operating or planned very large
spectroscopic and time-domain surveys, like SDSS, LAMOST, LSST, and future X-ray surveys.

\section{Candidate recoiling SMBHs identified by kinematic signatures}

\subsection{SDSSJ092712.65+294344.0, and X-ray follow-ups}

The quasar SDSSJ092712.65+294344.0 (SDSSJ0927+2943 hereafter; Komossa et al. 2008, 
KZL08 hereafter) at $z$=0.7 shows all the characteristic
optical signatures of a recoiling SMBH, which were predicted by Bonning et al. (2007):
Its broad emission lines are shifted by 2650 km s$^{-1}$ with respect to its
narrow emission lines, the broad lines are symmetric, the broad MgII line shows
the same shift as the broad Balmer lines, and the narrow emission lines lack
an ionization stratification as expected if the accreting SMBH is no longer at the center
of the system
(KZL08).{\footnote{SDSSJ0927+2943 also shows a second system of narrow emission lines
with unusual properties when compared with other known quasars, including
exceptionally broad Neon emission lines. The origin
of these lines is still being explored; the lower-ionization lines are too
narrow to have originally been bound to the recoiling SMBH (except in case
of projection effects), and their low degree of ionization is not
straightforward to understand (KZL08).  A possible reservoir of
narrow-line gas at the kick velocity is stellar mass loss, as a consequence
of stellar evolution of the stars bound to the recoiling SMBH (Komossa \& Merritt 2008a).}}
Its unique properties
make SDSSJ0927+2943 an excellent candidate for a recoiling SMBH.

Two alternative models have been considered in order
to explain some (but not all) of the unusual properties of this system;
a chance projection, within 1 arcsec, 
of one or {\em two} {\em intrinsically peculiar} AGN in a very
massive cluster of galaxies (KZL08, Shields et al. 2009a, Heckman et al. 2009),
and a close pre-merger binary SMBH (Dotti et al. 2009, Bogdanovi{\'c} et al. 2009).
However, a rich and massive cluster has not been detected in NIR and X-ray
imaging follow-up observations (Decarli et al. 2009, Komossa et al. in prep.).
Neither was the predicted orbital motion of an SMBH binary detected in spectroscopic
follow-ups (Shields et al. 2009a; see also Vivek et al. 2009).
This leaves us with the recoil scenario for 
SDSSJ0927+2943. This scenario is also consistent with the
recent measurement of an {\em off-set} between the QSO and the host galaxy
as traced by [OIII] emission (Vivek et al. 2009).

We have obtained an imaging observation of SDSSJ0927+2943 with the $Chandra$
X-ray observatory, in order
to measure more precisely its X-ray luminosity (than was possible with
a serendipituous off-axis ROSAT observation; KZL08), and to study the properties
of the field around SDSSJ0927+2943, including the search for a possible
massive cluster of galaxies. We detect point-like X-ray emission from the quasar
coincident with the optical position of SDSSJ0927+2943. A second X-ray source
is present at a distance of $\sim$17 arcsec from SDSSJ0927+2943.
This second source coincides with the object SDSSJ092713.8+294336 and contributed
approximately 70\% to the ROSAT X-ray emission from the region of SDSSJ0927+2943.
Luminous extended X-ray emission from a {\em rich} cluster, in the form predicted by
Heckman et al. (2009), is not present. The full results of the X-ray analysis will
be presented by Komossa et al. (in prep).

\subsection{E1821+643}
The well-known, luminous quasar E1821+643 ($z$ = 0.297) 
shows highly asymmetric broad Balmer lines which appear
different in direct and in polarized light, and are strongly shifted with respect to the
narrow lines. Based on their spectropolarimetric
observations, Robinson et al. (2010) favor a scenario where one component of the
BLR is bound to a recoiling black hole, which is moving at 2100 km/s relative to
its host galaxy.     
A second broad-line system is shifted by only 470 km/s, and its nature is currently
unclear. If still related to recoil, in form of a marginally bound or unbound component
of the BLR, the system is young, and Robinson et al. then
estimate an age of $\sim$10$^{4}$ years.

\subsection{SDSSJ105041.35+345631.3}
Shields et al. (2009) selected the quasar SDSSJ105041.35+345631.3 at $z$ = 0.272 from the SDSS
because of its large kinematic shift of the BLR of, in this case, 3500 km/s relative
to the narrow emission lines. A projection effect is considered unlikely, as is a
binary SMBH because of the lack of detectable orbital motion. While Shields et al.
do not rule out an extreme case of a recoiling SMBH, they conclude that several
aspects of the optical spectrum are best understood if this galaxy is an extreme case of
a ``double-peaked emitter''.

\section{Candidate recoiling SMBHs identified by spatial off-sets}

\subsection{CID-42}
The galaxy CID-42 (COSMOSJ1000+0206) at redshift $z$=0.359 was discovered
in the COSMOS survey (Elvis 2009), and caught attention due to its unusual morphology
with two apparent optical ``nuclei'' (Comerford et al. 2009) at 
a projected separation of 2 kpc, and an extended tidal tail.
Initially suspected to be a binary AGN (Comerford et al. 2009), it was
then re-interpreted as a candidate recoiling SMBH, or alternatively, an SMBH ejection following
3-body interaction in a triple SMBH system, by Civano et al. (2010). 
An HST image analysis has shown that the
north-western core is slightly extended though compact, and consistent with being the 
nucleus of the galaxy, while only the south-eastern bright source is point-like
(Civano et al 2010).     
The optical spectrum of CID 42 shows a kinematic shift of 1200 km/s 
between the BLR and the major
narrow-line component, and 
extra faint narrow H$\beta$ emission at the same redshift as the broad lines
(Civano et al. 2010), and perhaps further faint narrow-line emission shifted
by $\sim$150 km/s (Comerford et al. 2009). As such, the spectrum shares similarities
with that of the recoil candidate  SDSSJ092712.65+294344.0 
(Komossa et al. 2008, Shields et al. 2009a).  
Another remarkable feature, not yet well understood, is the
presence of a strong redshifted broad iron line with a P-Cygni profile,
variable over four years, of high column density and highly ionized (Civano et al. 2010).
Follow-up optical and X-ray observations are currently underway
(Civano et al. 2012, in prep.). Their new high-resolution Chandra data show the presence 
of only one X-ray emitting object which coincides with the position
of the south-eastern optical source, supporting the recoil scenario 
(Civano et al. 2012, in prep.).

\subsection{M87} 
M87 is a nearby massive galaxy with a prominent radio jet. 
The photo-center
of the host galaxy is off-set by 7 pc from the nuclear point source (i.e., presumably the 
location of the SMBH) (Batcheldor et al. 2010). 
The displacement is in the direction of the counterjet. Among several scenarios 
(acceleration by a jet,
presence of massive perturbers, binary orbital motion) considered, 
Batcheldor et al. (2010) 
favor GW recoil as the most plausible. The observed off-set can then be
explained either by a  moderate kick 1 Myr ago, or residual small-amplitude
oscillations of a large recoil which happened $<$ 1 Gyr ago.

\subsection{CXOJ122518.6+144545}
Jonker et al. (2010) reported the detection of an unusual off-nuclear X-ray source,
at a projected separation of 3 kpc from the core of the galaxy SDSSJ122518.86+144547.7  
at $z$=0.045.
CXOJ122518.6+144545 is X-ray luminous, has a bright optical counterpart and  
properties unlike those of other off-nuclear X-ray sources which were found in large numbers
with $Chandra$. 
The authors offer three explanations of CXOJ122518.6+144545: 
a supernova of type IIn, an ultraluminous X-ray source with 
an unusually bright optical counterpart, or a recoiling SMBH. 
Bellovary et al. (2010) further discuss the possibility of a wandering SMBH in the
galaxy halo, produced by stripping of a satellite which merged with the primary galaxy.

\subsection{ESO 1327-2041}
The nearby galaxy ESO 1327-2041 ($z$ = 0.018) shows a complex morphology
indicative of a recent merger. $HST$ imaging has revealed the presence of a
compact source embedded in an extended ``stellar stream'' (Keeney et al. 2010;
their Fig. 1), at a redshift similar to the core of ESO 1327-2041, and at a 
projected separation of 15 kpc. 
Keeney et al. discuss several possible interpretations of this compact object,
and propose that it is the actual nucleus of the galaxy, ejected as a consequence 
of either tidal interaction between two galaxies or gravitational wave
recoil following a past merger. 

\section{Implications of recoil oscillations for unified models of AGN}

There are potentially far-reaching consequences of SMBH recoil
for unified models of AGN.
Spatial oscillations of the SMBHs about the
cores of their host galaxies imply that the SMBHs spend a
significant fraction of time {\em off-nucleus}, at scales beyond that of the molecular
obscuring torus. An intrinsically {\em obscured} quasar of {\em type 2} with its BLR hidden
by the torus will therefore appear as {\em unabsorbed, type 1} quasar during
the recoil oscillations, when moving beyond the torus scale.
Assuming reasonable
distributions of recoil velocities,
Komossa \& Merritt (2008b) have
computed the off-core timescale of (intrinsically type-2)
quasars{\footnote{These
calculations are based on models of Gualandris \& Merritt (2008),
which did not include gas. Recent simulations
of recoil oscillations in a gaseous disk show, that oscillation
timescales can either increase or decrease (Sijacki et al. 2010) with respect to
the gas-free case.}}.
It was shown that roughly 50\% of all
major mergers result in a SMBH being displaced beyond the torus
for a time of $10^{7.5}$ yr or more. This is an interesting
number, because it is
comparable to quasar activity time
scales.  Since {\em major}
mergers (i.e., quasars) are most strongly affected by gravitational wave recoil,
our results imply a deficiency of luminous type 2 {\em quasars}
in comparison to low-luminosity {\em Seyfert} 2 galaxies, as indeed observed
(e.g., Hasinger 2008). These may
therefore naturally explain the long-standing puzzle, why few
absorbed type 2 quasars exist at high luminosities; it would
be these which are affected by the recoil oscillations, therefore
appearing as type 1 rather than type 2 for a significant
fraction of their lifetime (Komossa \& Merritt 2008b).

Recoil oscillations further imply the presence of a fraction of
quasars which lack a hot dust component, because the dusty torus
is only illuminated from a distance. Such ``hot-dust-poor'' quasars
have indeed been observed (e.g., Hao et al. 2011).

Recoil oscillations also have a number of other
observable consequences related to AGN.
For instance, they will affect the X-ray background and its modeling since
a fraction of sources will be unobscured at any given time.
In particular, small amplitude oscillations of the order the torus size
will affect the ratio of Compton-thin to Compton-thick sources, and
could lead to measurable variability in the absorption and extinction
of AGN spectra once the recoiling SMBH passes the individual clouds
making up the torus (Komossa \& Merritt 2008b).

\section{Astrophysical implications and future observations}

The kicks and superkicks predicted by recent numerical relativity simulations
of coalescing SMBHs have stimulated  
an active new field of research. Electromagnetic signatures of recoiling
SMBHs are being predicted, several candidates have emergerd in large data bases,
and astrophysical implications of the kicks
are still being explored.  The fact that SMBHs will not always reside at the very
cores of their host galaxies, or may even be ejected completely, has many
potential implications for the topics discussed in this book;  
for galaxy and SMBH assembly and galaxy-SMBH (co-)evolution, core structures
in early type galaxies, the scatter in the host galaxy -- SMBH scaling relations,
the statistics of obscured quasars, and the redshift-dependence of gravitational
wave signals  
(e.g., Haiman 2004, Boylan-Kolchin et al. 2004, Libeskind et al. 2006, 
Schnittman 2007, Sesana 2007, Tanaka \& Haiman 2009, Volonteri et al. 2010, 
Micic et al. 2011). 

It is therefore important to identify more
candidate recoiling SMBHs through
observations. Promising future searches would include:
(1) emission-line signatures in large spectroscopic data bases such as SDSS or LAMOST;
(2) recoil flares from accretion disks and stellar tidal
disruptions in large-scale surveys like Pan-STARRS, LSST and especially in X-rays,
and (3) the characteristic, large stellar velocity dispersions of HCSSs in
spectroscopic follow-ups of ongoing imaging surveys of nearby clusters of galaxies.

Detecting recoiling SMBHs in large numbers will open up a new window on measuring
galaxy merger histories and kick amplitude distributions, and testing predictions
of numerical relativity.



\begin{thebibliography}{}

\bibitem[Baker et al. (2008)]{baker08}
{Baker, J.G., et al.} 2008,
\textit{ApJ} (Letters), 682, L29

\bibitem[Batcheldor et al. (2010)]{Bat10}
{Batcheldor, D., et al.} 2010,
\textit{ApJ} (Letters), 717, L6

\bibitem[Begelman et al. (1980)]{begelman80}
{Begeglman, M.C., Blandford, R.D., Rees, M.J.} 1980,
\textit{Nature}, 287, 307 

\bibitem[Bekenstein (1973)]{bekenstein73}
{Bekenstein, J.D.} 1973,
\textit{ApJ}, 183, 657  

\bibitem[Bellovary et al. (2010)]{Bel10}
{Bellovary, J., et al.} 2010,
\textit{ApJ} (Letters), 721, L148 

\bibitem[Bloom et al. (2011)]{Bloom11}
{Bloom, J., et al.} 2011,
\textit{Science}, 333, 203   

\bibitem[Bonning et al. (2007)]{Bo2007}
{Bonning, E.W., Shields, G.A., Salviander, S.} 2007,
\textit{ApJ} (Letters), 666, L13

\bibitem[Bogdanovic et al. (2007)]{bogda07}
{Bogdanovi{\'c}, T., Reynolds, C.S., Miller, C.} 2007,
\textit{ApJ} (Letters), 661, L147

\bibitem[Bogdanovic et al. (2009)]{bogda09}
{Bogdanovi{\'c}, T., et al.} 2009,
\textit{ApJ}, 697, L288

\bibitem[Boylan-Kolchin et al. (2004)]{Boylan-K04}
{Boylan-Kolchin, M., et al. } 2004,
\textit{ApJ} (Letters), 613, L37

\bibitem[Blecha and Loeb (2008)]{BL2008}
{Blecha, L., Loeb, A.} 2008,
\textit{MNRAS}, 390, 1311  

\bibitem[Bruegmann et al. (2008)]{bruegmannetal08}
{Br\"ugmann, B., et al.} 2008,
\textit{Phys. Rev. D.}, 77, 124047

\bibitem[Campanelli et al. (2007)]{campanelli07a}
{Campanelli, M., et al.} 2007,
\textit{ApJ} (Letters), 659, L5

\bibitem[Campanelli et al. (2009)]{campanelli09}
{Campanelli, M., et al.} 2009,
\textit{Phys. Rev. D.}, 79, 084010

\bibitem[Centrella et al. (2010)]{cen2010}
{Centrella, J., et al.} 2010,
\textit{Rev. Mod. Phys.}, 82, 3069

\bibitem[Chen et al. (2009)]{chen09}
{Chen, X., et al.} 2009,
\textit{ApJ} (Letters), 697, L149 

\bibitem[Civano et al. (2010)]{civano2010}
{Civano, F., et al.} 2010,
\textit{ApJ}, 717, 209 

\bibitem[Comerford et al. (2009)]{comerford2009}
{Comerford, J.M., et al.} 2009,
\textit{ApJ} (Letters), 702, L82 

\bibitem[Corrales et al. (2010)]{Corrales10}
{Corrales, L.R., Haiman, Z., MacFadyen, A.} 2010,
\textit{MNRAS}, 404, 947  

\bibitem[Dain et al. (2008)]{dain08}
{Dain, S., Lousto, C., Zlochower, Y.} 2008,
\textit{Phys. Rev. D.}, 78, 024039

\bibitem[Decarli et al. 2009]{Decarli09}
{Decarli, R., Reynolds, M.T., Dotti, M.} 2009,
\textit{MNRAS}, 397, 458

\bibitem[de la Fuente Marcos (2008)]{delafuente08}
{de la Fuente Marcos, R., de la Fuente Marcos, C.} 2008,
\textit{ApJ} (Letters), 677, L47

\bibitem[Devecchi et al. (2009)]{Deve09}
{Devecchi, B., et al.} 2009,
\textit{MNRAS}, 394, 633

\bibitem[Dotti et al. (2009)]{dotti09}
{Dotti, M., et al.} 2009,
\textit{MNRAS}, 398, L73

\bibitem[Dotti et al. (2010)]{dot10}
{Dotti, M., et al.} 2010,
\textit{MNRAS}, 402, 682  

\bibitem[Elvis (2009)]{elvis09}
{Elvis, M.} 2009,
\textit{BAAS}, 41, 708

\bibitem[Eracleous et al. (2011)]{era11}
{Eracleous, M., et al.} 2011,
arXiv:1106.2952

\bibitem[Fujita (2009)]{fujita09}
{Fujita, Y.} 2009,
\textit{ApJ}, 691, 1050

\bibitem[Gonzales et al. (2007)]{gonzalesetal07b}
{Gonz{\'a}les, J.A., et al.} 2007,
\textit{Phys. Rev. Lett.}, 98, 231101

\bibitem[Gonzales et al. (2009)]{gonzalesetal09}
{Gonz{\'a}les, J.A., Sperhake, U., Br\"ugmann, B.} 2009,
\textit{Phys. Rev. D.}, 79, 124006

\bibitem[Gualandris et al. (2008)]{gua08}
{Gualandris, A., Merritt, D.} 2008,
\textit{ApJ}, 678, 780

\bibitem[Haiman (2004)]{haiman04}
{Haiman, Z.} 2004,
\textit{ApJ}, 613, 36

\bibitem[Hao et al. (2011)]{hao11}
{Hao, et al.} 2011,
\textit{ApJ}, 733, 108

\bibitem[Hasinger (2008)]{hasinger08}
{Hasinger, G.} 2008,
\textit{A\&A}, 490, 905

\bibitem[Healy et al. (2009)]{healy09}
{Healy, J., et al.} 2009,
\textit{Phys. Rev. Lett.}, 102, 041101

\bibitem[Heckman et al. (2009)]{heck09}
{Heckman, T. et al.} 2009,
\textit{ApJ}, 695, 363

\bibitem[Herrmann et al. (2007b)]{herrmann07b}
{Herrmann, F., et al.} 2007,
\textit{Phys. Rev. D}, 76, 084032

\bibitem[Holley-Bockelmann et al. (2008)]{holley08}
{Holley-Bockelmann, K., et al.} 2008,
\textit{ApJ}, 686, 829  

\bibitem[Jonker et al. (2010)]{jonker10}
{Jonker, P.G., et al.} 2010,
\textit{MNRAS}, 407, 645  

\bibitem[Keeney et al. (2010)]{keeney10}
{Keeney, B.A., et al.} 2010,
\textit{AJ}, 141, 66 

\bibitem[Kesden et al (2010)]{kesden10}
{Kesden, M., Sperhake, U., Berti, E.} 2010,
\textit{ApJ}, 715, 1006  

\bibitem[Komossa (2006)]{komossa06}
{Komossa, S.} 2006,
\textit{MmSAI} 77, 733 

\bibitem[Komossa and Bade (1999)]{koba99}
{Komossa, S., Bade, N.} 1999,
\textit{A\&A} 343, 775

\bibitem[Komossa and Merritt (2008a)]{KM2008}
{Komossa, S., Merritt, D.} 2008a,
\textit{ApJ} (Letters), 683, L21 (KM08)

\bibitem[Komossa and Merritt (2008b)]{KM2008b}
{Komossa, S., Merritt, D.} 2008b,
\textit{ApJ} (Letters), 689, L89

\bibitem[Komossa et al. (2008)]{KZL08}
{Komossa, S., Zhou, H., Lu, H.} 2008,
\textit{ApJ} (Letters), 678, L81 (KZL08)

\bibitem[Kornreich (2008)]{kolovelace08}
{Kornreich,  D.A., Lovelace, R.V.E.} 2008,
\textit{ApJ}, 681, 104

\bibitem[Le Tiec et al. (2010)]{LeTiec10}
{Le Tiec, A., Blanchet, L., Will, C.M.} 2010,
\textit{Class. Quant. Grav.}, 27, 012001  

\bibitem[Libeskind et al. (2006)]{libeskind06}
{Libeskind, N.I., et al.} 2006,
\textit{MNRAS}, 368, 1381

\bibitem[Lippai et al. (2008)]{lip08}
{Lippai, Z., et al.} 2008,
\textit{ApJ} (Letters), 676, L5

\bibitem[Liu et al. (2003)]{liu03}
{Liu, F. et al.} 2003,
\textit{MNRAS}, 340, 411

\bibitem[Liu et al. (2011)]{liu11}
{Liu, F. et al.} 2011, \textit{ApJ}, in press, 
arXiv:1112.1081 

\bibitem[Loeb (2007)]{loeb07}
{Loeb, A.} 2007,
\textit{Phys. Rev. Lett.}, 99, 041103

\bibitem[Lousto et al. (2010)]{loustoetal10}
{Lousto, C., et al.} 2010,
\textit{Class. Quant. Grav.}, 27, 114006

\bibitem[Lousto and Zlochower (2009)]{lz09}
{Lousto, C., Zlochower, Y.} 2009,
\textit{Phys. Rev. D.}, 79, 064018

\bibitem[Lousto and Zlochower (2011a)]{LoustoZ11a}
{Lousto, C., Zlochower, Y.} 2011a
\textit{Phys. Rev. D.}, 83, 024003 

\bibitem[Lousto and Zlochower (2011b)]{LZl11b}
{Lousto, C., Zlochower, Y.} 2011b, arXiv:1108.2009

\bibitem[Madau et al. (2004)]{madau04}
{Madau, P., et al.} 2004,
\textit{ApJ}, 604, 484

\bibitem[Madau and Quataert (2004)]{madauquataert04}
{Madau, P., \& Quataert, E.} 2004,
\textit{ApJ} (Letters), 606, L17

\bibitem[Megevand et al. (2009)]{Megevand09}
{Megevand, M., et al.} 2009,
\textit{Phys. Rev. D.}, 80, 024012

\bibitem[Merritt et al. (2004)]{merritt04}
{Merritt, D., et al. 2004},
\textit{ApJ} (Letters), 607, L9

\bibitem[Merritt et al. (2006)]{merritt06}
{Merritt, D., et al.} 2006,
\textit{MNRAS}, 367, 1746

\bibitem[Merritt et al. (2009)]{MSK09}
{Merritt, D., Schnittman, J., Komossa, S.} 2009,
\textit{ApJ}, 699, 1690

\bibitem[Micic et al. (2011)]{mi11}
{Micic, M.., et al.} 2011,
\textit{MNRAS}, 414, 1127  


\bibitem[Miller and Matzner (2009)]{millermatzner09}
{Miller, S.H., Matzner, R.A.} 2009,
\textit{GReGr}, 41, 525

\bibitem[Milosavljevic and Phinney (2005)]{milos05}
{Milosavljevi{\'c}, M., Phinney, S.} 2005,
\textit{ApJ} (Letters), 622, L93

\bibitem[Natarajan and Armitage (1999)]{nata99}
{Natarajan, P., Armitage, P.J.} 1999,
\textit{MNRAS}, 309, 961

\bibitem[O'Leary and Loeb (2009)]{oleary09}
{O'Leary, R.M., Loeb, A.} 2009,
\textit{MNRAS}, 395, 781

\bibitem[O'Leary and Loeb (2011)]{oll11}
{O'Leary, R.M., Loeb, A.} 2011,
arXiv:1102.3695

\bibitem[Perego et al. (2009)]{Perego09}
{Perego, A., et al.} 2009,
\textit{MNRAS}, 399, 2249

\bibitem[Peterson (2007)]{peterson07}
{Peterson, B.M. 2007},
\textit{ASPC}, 373, 3

\bibitem[Ponce et al. (2011)]{ponce11}
{Ponce, M., et al.} 2011,
arXiv:1107.1711

\bibitem[Robinson et al. 2010 (2010)]{robinson10}
{Robinson, A., et al.} 2010,
\textit{ApJ} (Letters), 717, L122

\bibitem[Rossi et al. (2010)]{Rossi10}
{Rossi, E.M., et al.} 2010,
\textit{MNRAS}, 401, 2021 

\bibitem[Scheuer and Feiler (1996)]{Scheuer96}
{Scheuer, P.A.G., Feiler, R.} 1996,
\textit{MNRAS}, 282, 291

\bibitem[Schnittman (2007)]{schnittman07}
{Schnittman, J.D.} 2007,
\textit{ApJ} (Letters), 667, L133

\bibitem[Schnittman and Buonanno (2007)]{schnittbu07}
{Schnittman, J.D., Buonanno, A.} 2007,
\textit{ApJ} (Letters), 662, L63

\bibitem[Schnittman and Krolik (2008)]{schnittkro08}
{Schnittman, J.D., Krolik, J.H} 2008,
\textit{ApJ}, 684, 835

\bibitem[Shields and Bonning (2008)]{shibo08}
{Shields, G.A., Bonning, E.W.} 2008,
\textit{ApJ}, 682, 758

\bibitem[Shields et al. (2009a)]{shields09a}
{Shields, G.A., Bonning, E.W., Salviander, S.} 2009a
\textit{ApJ}, 696, 1367

\bibitem[Shields et al. (2009b)]{sh09b}
{Shields, G.A., et al.} 2009b
\textit{ApJ}, 707, 936

\bibitem[Sesana (2007)]{sesana07}
{Sesana, A.} 2007,
\textit{MNRAS}, 382, L6

\bibitem[Sijacki et al (2011)]{sijacki11}
{Sijacki, D., Springel, V., Haehnelt, M.G.} 2011,
\textit{MNRAS}, 414, 3656 

\bibitem[Sperhake et al. (2007)]{sperrhake11}
{Sperhake, U., et al.} 2011,
arXiv:1011.3281

\bibitem[Stone and Loeb (2011)]{stone11}
{Stone, N., Loeb, A.} 2011,
\textit{MNRAS}, 412, 75 

\bibitem[Tanaka and Haiman (2009)]{Tanaka09}
{Tanaka, T., Haiman, Z.} 2009,
\textit{ApJ}, 696, 1798

\bibitem[Tanaka et al. (2010)]{Tana10}
{Tanaka, T., Haiman, Z., Menou, K.} 2010,
\textit{AJ}, 140, 642

\bibitem[Vivek et al. (2009)]{Vivek09}
{Vivek, M., et al.} 2009,
\textit{MNRAS}, 400, L6

\bibitem[Volonteri et al. (2007)]{volonteri07}
{Volonteri, M., Sikora, M., Lasota, J.-P.} 2007,
\textit{ApJ}, 667, 704  

\bibitem[Volonteri et al. (2010)]{volonteri10}
{Volonteri, M., et al.} 2010,
\textit{MNRAS}, 404, 2143  

\bibitem[Volonteri and Madau (2008)]{volomadau08}
{Volonteri, M., Madau, P.} 2008,
\textit{ApJ} (Letters), 687, L57

\bibitem[Zanotti et al. (2011)]{zan11}
{Zanotti, O., et al.} 2011,
\textit{MNRAS}, 417, 2899  

\end{thebibliography}
\end{document}